# A/D Converter Architectures for Energy-Efficient Vision Processor

Li Du, Yilei Li

0572493@fudan.edu.cn

*Abstract-* **AI applications have emerged in current world. Among AI applications, computer-vision (CV) related applications have attracted high interest. Hardware implementation of CV processors necessitates a high performance but low-power image detector. The key to energy-efficiency work lies in analog-digital converting, where output of imaging detectors is transferred to digital domain and CV algorithms can be performed on data. In this paper, analog-digital converter architectures are compared, and an example ADC design is proposed which achieves both good performance and low power consumption.**

## I. INTRODUCTION

WITH the advent of artificial intelligence (AI), new applications have been proposed. In AI applications, CV applications have attracted especially attention [1-5], as CV applications can help people do a lot of routine works in daily life, such as face detection, pose estimation [2], image processing [3] etc.

The algorithms of AI have evolved at quick pace during previous a few years. In 2012, AlexNet [1] was proposed and it has been proved to be a large step for both deep learning and AI. With the success of deep learning, people are using it for all kinds of applications in daily life. However, hardware side of AI and deep learning only gets relatively less attention, though it is equally important.

## II. CV, VISION PROCESSOR AND ADC

CV algorithms need to be used in daily life, and embedded system is an ideal platform for daily uses (such as on mobile phone or wearable devices). However, common CV application algorithms need heavy computation load.

For example, hand-object interaction is commonly used in augmented reality applications. A new and efficient hand-object interaction algorithm that can provide hand segmentation from depth map has been proposed in [2]. In [2], bilateral filtering is used for data processing and excellent classification results can be achieved in proposed algorithm. However, the algorithm runs on a powerful server with GPU, which is not applicable for wearable devices such as HoloLens.

In order to be used in wearable devices, vision processors must be used. Compared with general-purpose GPU, vision processor is supposed to process CV tasks, and specially designed ASIC can be used as vision processor. Vision processor can provide energy-efficient operation.

A typical vision processor is composed of both analog and digital parts, as shown in Fig.1. The image sensor part senses image, and analog signal is generated by it. In order to perform CV algorithms in digital domain, analog signal must first be converted by analog-digital converter (ADC).

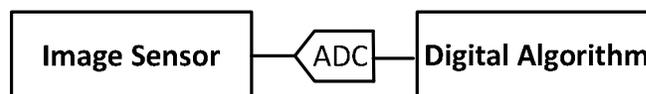

Fig. 1. System diagram of vision processor

ADC therefore plays a critical role in vision processor. First, it needs to provide accurate conversion between analog and digital. If ADC is poor in linearity or noise, the image from sensor can be distorted or corrupted. The distorted or corrupted image can significantly degrade the performance of CV algorithm.

Another key metric of ADC is power consumption. In order to improve energy efficiency of vision processor, ADC must consume low power.

## III. COMPARISON OF ADC ARCHITECTURES

CMOS technology has been evolving at rapid pace and can be used to implement many high precision, high speed circuits [6-7]. To achieve an efficient vision system, CMOS front end sensor array is also required to be low power, thus to request an efficient ADC as the front end sensing block to bridge the analog signal and digital signal processing world. In this section, we compare three most common ADC architectures [8].

The first category o ADC is sigma-delta ADC, as shown in Fig. 2. Sigma-delta (SD) ADC uses SD modulator, which turns oversampling into SNR enhancement. With SD ADC, very high resolution can be achieved. However, considering oversampling ratio, the real Nyquist rate of SD ADC is usually not high.

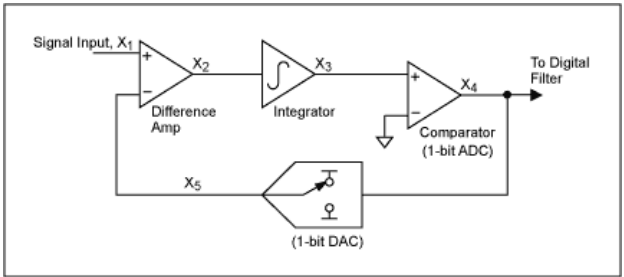

Fig. 2 SD ADC diagram

For image sensor applications, many sensors usually time-share one same ADC, and each ADC will do A/D conversion in a time-interleaving manner. Therefore one ADC needs to sample at moderate speed with low power. Though one SD ADC can serve one image sensor pixel at low power consumption, it would be consuming too much power if we implement many SD ADCs for image sensor pixels, or one fast SD ADC to time share.

The second category of ADC is flash ADC (Fig. 3). Flash ADC can achieve the highest sampling rate among all categories of ADCs, but its resolution is limited by mismatch among resistor ladder. Mismatch trimming (using laser) can be implemented to enhance resolution, but the cost is too high.

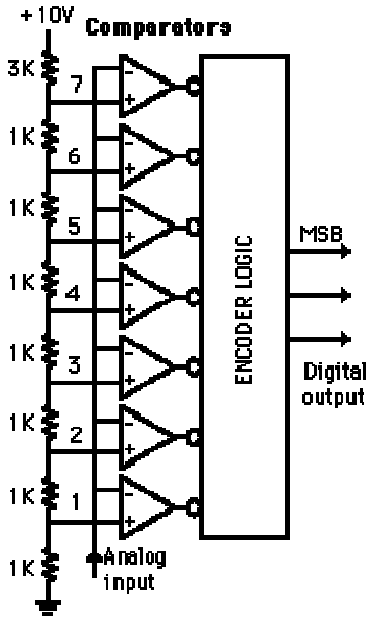

Fig. 3 Diagram of flash ADC.

The third architecture is pipelined ADC (Fig. 4). Pipelined structure, as one of the typical architectures has been widely implemented in the ADC design. A low power, middle-resolution (7~10 bit), middle speed (20MHz-200MHz) pipelined ADC has been used in many applications.

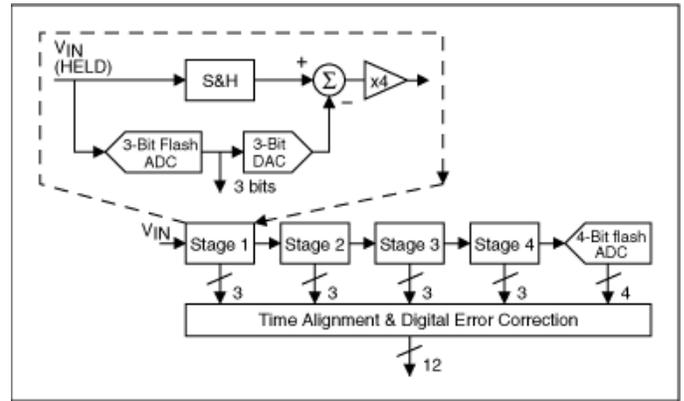

Fig. 4 Pipeline ADC

Pipeline ADC uses multiple stages, each stage will turn input analog signal into coarse digital code, amplify residue analog signal and send to the next stage. The main issues with pipeline ADC are its power consumption and conversion error. For power consumption, since each amplifier is used in each stage, power consumption can be high. For conversion error, mismatch between stages can introduce significant resolution degradation and nonlinearity of ADC.

Fortunately those two problems with pipeline ADC can be solved by design techniques. Various power reduction techniques have been developed for pipelined ADCs, such as gain calibration for the sample and hold amplifier, flash ADC-based MSB quantization, and removing the front-end sample/hold amplifiers [9].

For conversion error, calibration technique can be used to help. A group of calibration techniques have been developed to compensate the most significant MDAC gain error. In [10-11], a reference ADC was used to calibrate a single nonlinear MDAC by estimating its 3rd-order harmonic term. Complicated adaptation algorithms, such as least-mean-square (LMS), however, are needed for the estimation of MDAC parameters. In [12], a digital processor with 8.4 K gates was implemented in order to use the adaption algorithms and it consumes a large area. The LMS loop also leads to a trade-off between step size and convergence speed. Besides, the adaption-based calibration techniques are not scalable as technology improves.

Considering the pros and cons of typical ADCs, pipeline ADC is a good candidate for energy-efficient vision processors.

IV. EXAMPLE DESIGN OF PIPELINE ADC ARCHITECTURE

*A. System Level Architecture*

In order to minimize the power consumption and increase the speed as much as possible, an operational transconductance amplifier (OTA) sharing architecture for two series stages have been implemented. The fundamental benefit is the fact that for pipelined ADC, the OTA is only working in residue amplifier mode. During sampling mode, the OTA is relaxed, as Fig. 5 shows. Meanwhile, we notice that each following stage is half of a cycle delayed compared to the current stage. So when the first stage is in sampling mode, the second one is in residue amplifying mode, and versa vice. For this reason, the number of OTAs can be reduced by using one OTA for both stages, the single-ended architecture is shown Fig. 6.

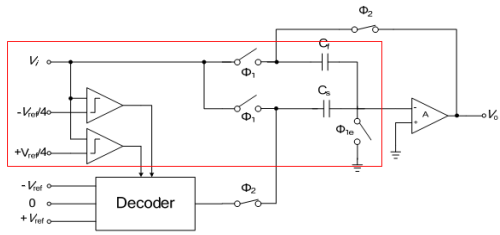

Fig.5 sampling mode of a pipeline stage

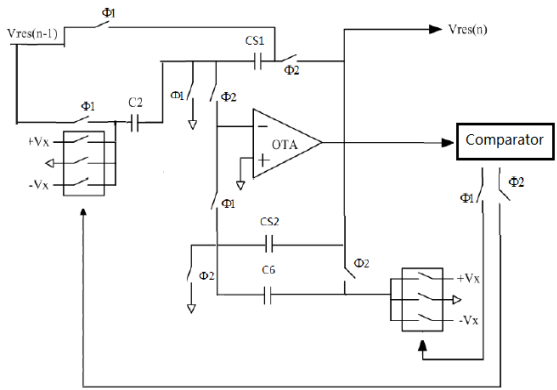

Fig. 6. Schematic of the sharing OTA

Based on this OTA sharing architecture, the whole system architecture can be drawn as Fig. 7. The ADC consists of a sample and hold amplifier in the front, and six 1.5bit stage with OTA sharing architecture in the middle. At last, a 2bit-flash stage is implemented to quantify the remaining 2 bits.

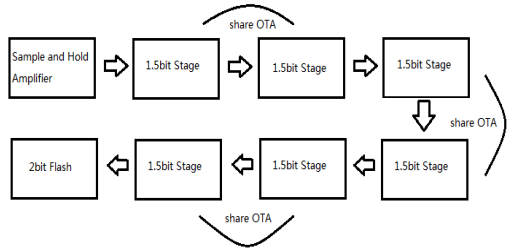

Fig. 7. Whole ADC architecture

### B. Sample and Hold Circuit

The switch capacitor-based sample and hold circuit is implemented as Fig. 8. In ck1, the circuit samples the signal voltage and stored into the sampling capacitor. In ck2, the OTA's input and output switches are turned on and the OTA forms a feedback loop. Thus the sampling capacitor's negative side voltage is set to 0 and the OTA generates the sampled voltage at its output.

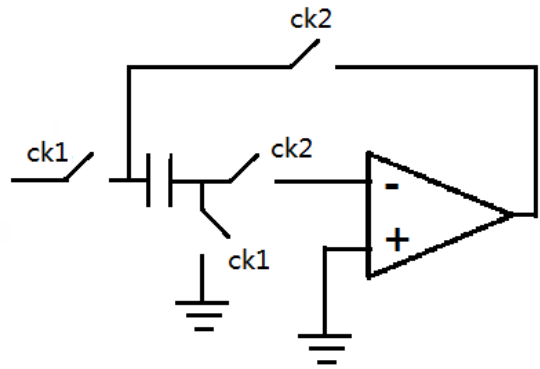

Fig.8 Sample and hold circuit

## V. ADC SIMULATION RESULT

### A. Stage Setup Simulation

In order to test the ADC's sample and hold performance, an input pulse from -600mV to 600mV has been injected into the ADC. In this case, the amplifier output needs to change from -600mV to 600mV in one cycle. The simulation result is listed as Tab.1.

| Stage | Ideal/mV | Simulation/mV | Error% |
|---|---|---|---|
| Vin | 600 | 600 | 0 |
| SHA | 600 | 599.7 | 0.05% |
| Stage1 | 599.7 | 599.2 | 0.08% |
| Stage2 | 599.2 | 598.5 | 0.11% |
| Stage3 | 598.5 | 596.3 | 0.36% |
| Stage4 | 596.3 | 593.9 | 0.4% |
| Stage5 | 593.9 | 587.4 | 1.1% |
| Stage6 | 587.4 | 575.6 | 2% |

Tab.1 Stage setup time simulation

As shown in the Tab.1, the input signal has been setup accurately (error<1%) till $5^{th}$ stage. The $5^{th}$ stage and $6^{th}$ stage have relative large setup error due to the accumulation of the errors from previous stages. However, at that stage, the requirement of the accuracy is relative low because most of the MSB have already been quantified.

### B. Linearity Simulation

To measure the INL and differential nonlinearity (DNL) of the ADC, a very slow ramp signal has been injected to the ADC input. The measured INL and DNL is shown Fig. 9. Based on the curve, maximum INL is 0.35LSB happened at code 224 and maximum DNL is 0.24LSB happened at code 96.

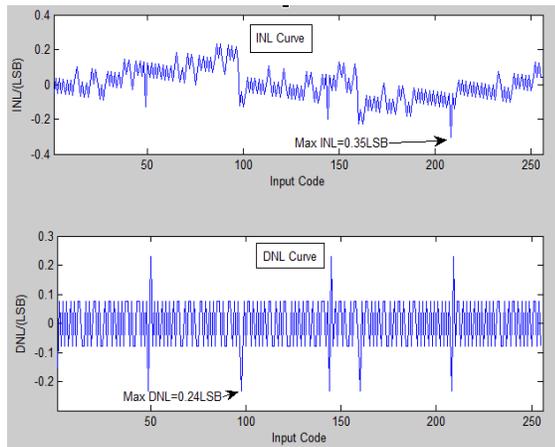

Fig.9 ADC INL DNL measurement

*C. Top Level Performance Simulation*

To verify the ADC's top level performance, the ADC was running at a sampling frequency of 166.6MHz. By running at such a high sampling frequency, one ADC can serve a huge number of image sensor pixels for vision processor. A 10.417MHz sinusoidal signal was injected into the ADC input. The measured output spectrum is shown in Fig.10. Based on this, the calculated signal-to-noise and distortion ratio (SNDR) is 45.9dB, Spurious-Free Dynamic Range (SFDR) is 50dB, and Effective number of bits (ENOB) is equal to 7.33 bit. This is enough for standard RGB format vision capture.

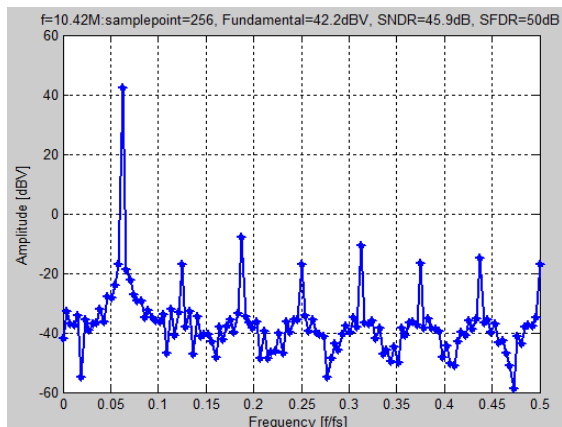

Fig.10 Spectrum of the ADC output

## VI. CONCLUSION

In this paper, we compared three typical categories of ADC for vision processor applications. Pipeline ADC is a good candidate for vision processor as it has fast sampling rate, decent resolution while consuming reasonable power. As an example design, an 8bit 166MS/s 38.9mW pipeline ADC with OTA sharing architecture is proposed. Simulation results show that the ADC has INL of 0.35LSB and DNL of 0.24LSB. When the input signal is 10.42MHz and sampling frequency is 166MS/s, it achieves an SNDR of 45.9dB, SFDR of 50dB and an ENOB of 7.33bit. This is a good example design for vision processors.